# Meeting Laboratory Course Learning Goals Remotely via Custom Home Experiment Kits

Daniel Howard, MSEE, Oglethorpe University, Division of Natural Sciences

Mariel Meier, PhD, Oglethorpe University, Division of Natural Sciences

*Abstract*— **In this paper, the use of cost-effective, custom home experiment kits for remote teaching of first-semester introductory physics labs is described. The kit experiments were designed to match the existing onsite lab experiment learning goals and the general laboratory course learning goals in remote teaching. Additionally, a revised group project approach optimized for remote use and that leverages the kits was developed and employed. Student survey results at the end of the Summer 2020 semester indicate that the critical learning goals were met, student satisfaction with the remote lab was maintained, and successful collaboration via video-conferencing breakout rooms was achieved.**

## I. Introduction

Following the emergency lockdown of the COVID-19 pandemic, introductory physics lab instructors had to adapt to remote delivery of their labs with limited success as recently reported by Fox et al. [1]. At Oglethorpe University, introductory physics labs that service both algebra- and calculus-based physics courses were transformed to an online format for the second half of the spring 2020 semester. The initial method used for emergency remote teaching of physics labs was live, interactive, instructor video demos of the labs, followed by students getting data from the instructor to analyze. The only formal modification of the course structure was that students who had not already collected the data for their end-of-semester group projects were allowed to convert their group project into a research paper on the same measurement and/or topic they had previously chosen via a literature review.

Several challenges, many of which matched those seen by Fox et al, were observed, including:

- Significant numbers of students appeared to 'tune out';
- Students performed poorly on uncertainty estimation and calculations, even though sources were provided;
- Most group experiment projects delivered far less than originally proposed, and the ones that were converted into research studies had far less collaboration; and
- The desire to 'dive deeper' into experiments evaporated.

Against this backdrop and the specter of long-term lock-downs, home experiment kits were sought to keep students engaged and achieve all critical learning goals. But the cost of commercial kits was prohibitive: over $200 per student, often with additional licensing/content subscription fees. The physics faculty and science division chair were all in agreement that adding to the existing lab fees was undesirable. Plus, many of the commercially-available home kits did not match the specific learning goals of the existing university experiment labs. An alternative approach of using only household equipment would not have permitted all experiments to be done and may have created equity issues for students. A custom kit that met all learning goals and financial constraints was deemed to be the best solution.

The structure of this paper is as follows: first, the institution, course and specific course learning goals are described. We then present the kit design requirements, fabrication, testing and modifications to improve robustness and adapt to supply chain issues. Details of how the kits were specifically leveraged for the group experiment projects in order to achieve course learning goals is then discussed. Results from a student survey presented at the end of the course are then described which detail the achievement of critical learning goals. Finally, the results are compared to the recent findings of Fox et al. and important differences noted, the most interesting of which is that the present work concludes that it is possible to replicate the onsite lab experience critical learning goals via customized home experiment kits and a new group project process.

## II. Institutional Context

Oglethorpe University is a regional PUI serving 1,385 students outside of Atlanta, GA. More than half of the students live on campus during a traditional semester; many that commute are traveling from their family homes. Two of the most popular majors on campus are Biology and the Dual-Degree Engineering program. As a result, in a typical fall or spring semester, enrollment in all sections of the introductory physics labs is between 50-90 students.

The onsite introductory physics lab at Oglethorpe meets for three hours one day per week and services students in both the algebra- and calculus- based courses. Lab activities alternate between traditional equipment-based labs focusing on developing experimental skills, which include a full lab report write up, and simulation-based "discovery" labs, which reinforce fundamental physics concepts. In addition, students work in small groups to complete an independent measurement project which they present at a university-wide poster session at the end of the semester. The following course learning goals guide the development of all laboratory activities:

LG-1. Demonstrate experimental design and analysis of data;

LG-2. Demonstrate familiarity with basic tools like spreadsheet analysis, automated and manual timing, digital multimeter, etc.;

LG-3. Understand the nature of scientific measurements (repeatability, uncertainty, bias, and precision);

LG-4. Understand how to define criteria for suitable evidence;

LG-5. Develop practical skills in running experiments/trials, problem-solving and trouble-shooting of experiments;

LG-6. Develop scientific habits of mind (critical thinking);

LG-7. Understand how modeling is used to explore equations and concepts;

LG-8. Construct knowledge and a deeper understanding of physics via direct experience;

LG-9. Learn how to analyze and visualize data in several ways;

LG-10. Evaluate results and analyze implications;

LG-11. Develop technical writing and presentation skills to communicate scientific ideas;

LG-12. Learn how to establish research goals; and

LG-13. Learn how to design, evaluate feasibility, and improve custom scientific experiments.

In addition to the traditional fall and spring semesters, introductory physics courses are offered during abbreviated four-week summer terms. This necessitated the immediate development of the remote lab kits for the Summer 2020 course offerings, the first of which enrolled 24 students.

### III. KIT REQUIREMENTS

Since students were not going to be supervised in person, safety was paramount in the kit design, as well as the kits being robust both for ease of use but also to eliminate any liability concerns. Learning requirements were to match onsite experiments as much as possible to replicate all activity-specific learning goals and provide opportunities to experimentally explore physics concepts covered in the lecture class.

To ensure approval by university leadership, the kits also had to be low cost, both in bill of materials (BOM) and in shipping (and thus low weight), and use readily-available parts that could be delivered to the instructor within a few weeks, in time to fabricate the kits and inexpensively ship them to students who were not local. To further reduce costs, a plan for local students to come and pick up the kits was developed that met university COVID-19 guidelines and requirements.

The final kit design requirements were that kit components should be usable for multiple experiments to minimize kit weight, the components should be designed to minimize setup time by students, and the design should make it possible for students to run entire labs by themselves. Note that in the survey results discussed in Section VI, some students actually reported developing better time management skills from having to set up experiments before the synchronous lab sessions, and also greater satisfaction from doing the entire process themselves. Thus, some amount of setup time may actually be a good thing.

### IV. DESIGN, DEVELOPMENT, FABRICATION, AND DEPLOYMENT OF THE KITS

The specific onsite experiments to replicate were:

- Measuring g: measure parabolic displacement vs. time via picket fence drop with PASCO photogate, and measure pendulum period vs. length and launch angle;

- Newton's 2nd Law: measure acceleration of a cart plus load connected to various hanging weights via a pulley with vanes and a PASCO photogate;

- Circular motion and centripetal force: match centripetal force from rotation frequency to spring restoring force and measure spring constant;

- Rotation and moment of inertia (MoI): measure MoI of apparatus before and after loading with washers to get washer MoI and compare to parallel axis theorem; and

- Collisions in 2D: validate conservation of momentum before and after collision of steel and glass balls using a curved ramp for both head-on and glancing collisions.

To replace photogates and custom tablets used for timing and other measurements in the onsite lab, the PHYPHOX smartphone app developed at a German university and available from https://phyphox.org/ was selected. In addition to an acoustic stopwatch that would allow students to time events verbally and use their hands for running experiments alone, the app includes many physics-related tools for measurements and a plethora of YouTube videos on how to use it. It has for example, an audio function generator, frequency meter, magnetometer, accelerometer, and other tools, many of which are planned for use in the second semester remote lab. Finally, it is free open source software (FOSS), so there is a reasonable chance that new capabilities will be added as smart phones further evolve.

Next, the labs were reviewed to identify parts that could be replaced with readily available items. To replace precision masses used in many of the experiments, three sizes of six washers each from local home improvement centers were used. A large quantity of each size was procured, weighed, and the average mass of each size were provided to students with the kit specifications. To replace the clamps used onsite, and to allow students creative license in modifying experiments, a roll of blue painter's tape was included. While regular masking tape is cheaper, painter's tape was specified so it would not leave sticky marks on students' desks and tables. Finally, for general use, a meter stick, plastic protractor, paper clips, a flexible plastic ruler with fuller (to use as a steel ball ramp), and 3 m of both nylon and cotton string were included. A steel ball and glass marble were specified since they are readily obtained, as were 3 sheets of carbon paper used in the onsite collision lab.

Next the onsite experiments were reviewed to determine components required for more complex experiments, in particular those that had custom apparati. The pendulum period vs. length and launch angle experiment was already possible with the above parts. Experiments that can be replicated with low cost hardware included Newton's 2nd Law, which could be done with a toy car, pulley block and the PHYPHOX acoustic stopwatch app. The rotation and moment of inertia lab had a custom apparatus that could be scaled down and fabricated from inexpensive lumber, a wooden dowel, a small pulley, and a lazy-susan rotating base. The steel curved ramp in the collisions in 2D lab was replaced with the notion that a flexible plastic ruler with a fuller down the middle could be bowed, held in place with a string (like an archery bow), and taped to two pieces of cardboard to make an inexpensive yet stable ramp for controlling the velocity of the steel ball for experiments. To further reduce variation for students, a rail which was a piece of U-channel aluminum from the local home improvement store was used. Initial tests with the setup verified that the steel ball on this low-cost ramp had a repeatable and reliable velocity. To make it easy for student to 'string' the plastic ruler, 4 holes at each corner were drilled.

Two experiments required a complete redesign for the home kits: the picket fence drop with photogate for measuring g, and the rotating spring apparatus with DC motor for centripetal force measurement and determination of spring constant. For the picket fence drop, an acoustic version of a photogate using the PHYPHOX audio function generator and a laptop's microphone was attempted using a large cardboard picket fence. While it was possible to capture an audio trace and determine the precise times at which the openings in the cardboard picket fence passed between the phone and the laptop mic (using Audacity, a FOSS audio editor), the peak-to-average ratio of the trace was so low and the sensitivity to background noise and reflections so high that it was deemed not viable for students to perform the experiment exactly as the onsite version.

Another approach to replicate the measuring g experiment was to take a video of the ball trajectory against a piece of cardboard with a grid on it, but this required a slo-mo capability not found on many students' phones, and MPEG compression motion prediction can also foil accurate displacement tracking. An attempt was also made to use the smartphone accelerometer, but multiple issues and lack of 100% support prevented its use.

Measuring g via varying height of steel ball drops works well with the PHYPHOX acoustic stopwatch app, but loses the learning goal in the original lab of starting with displacement vs. time, calculating the instantaneous velocity vs. time midpoints via the Merton method, and finally determining acceleration as slope of the velocity vs. time curve.

Since a solution for the collisions in 2D experiment was now in place, an experiment was sought that might leverage the setup and still meet the activity learning goals. Mentzer's modification [2] of the Harvard lab for measuring g via ballistic ball trajectory was tested, and instead of using a long aluminum rail for measuring the ball velocity prior to flying through the air, the free fall drop timed with acoustic stopwatch app was added to the beginning of the procedure by instructing students to drop the ball from the same height as the aluminum rail. Instead of a meter stick with tracing paper, carbon paper, and backdrop to mark the ball's trajectory, the 40" long shipping box was used with the carbon paper only taped to it. With this approach, it was possible for the instructor to reliably get within 5% of the actual value of g, and by adding the conventional pendulum experiment, 3 different methods of measuring g could be used by the students and compared, in keeping with the remaining goals of the original experiment. Figure 1 shows the setup and the actual ramp made out of the plastic ruler.

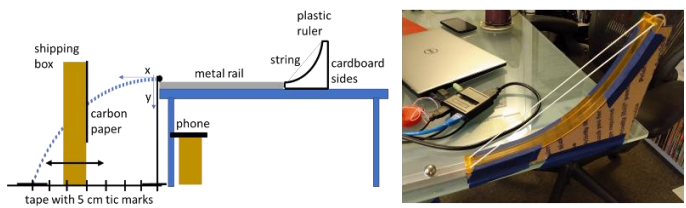

**Figure 1. Ballistic trajectory setup for measuring g (left) and actual curved ramp and rail from kit parts.**

The MoI apparatus is shown in Figure 2. The kit version (bottom) is a smaller and lighter version of the onsite lab apparatus (top). The wooden drum is cutout from 3" hole-borer bit, the base is 1x2 whitewood about 15" long, and a 6d nail is bent into zig-zag shape to hold the red plastic pulley. The rotating mount is a 2" lazy-susan base, and screws fasten the 4" long pulley block to the base, and are removable for the Newton's 2nd Law experiment described later.

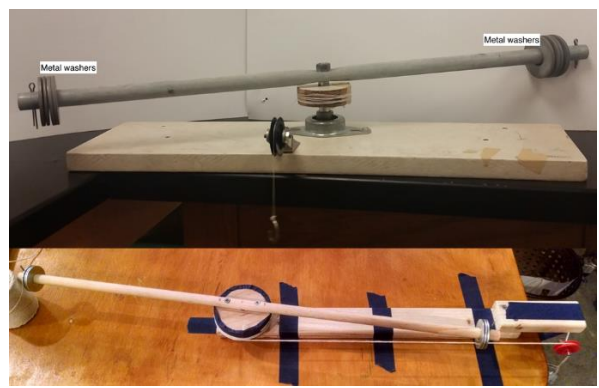

**Figure 2. Original (top) and kit version of MoI apparatus.**

Knowing that a wooden dowel would be used for the MoI experiment, it was also decided to use this for the pendulum experiment as a replacement for the usual clamp stands, as shown in Figure 3.

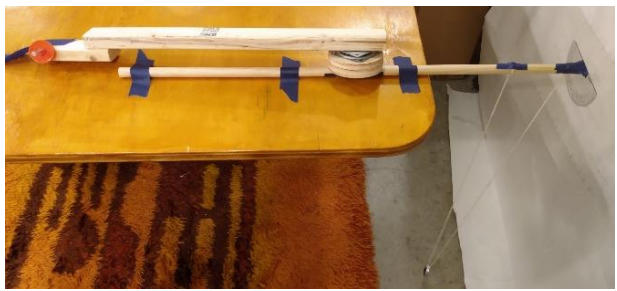

**Figure 3. MoI apparatus used for pendulum experiment.**

For the experiment on Newton's 2nd Law, the precision cart plus smart pulley w/photogate was replaced with a HotWheels™ car plus 6 large washers taped to the car as load and the pulley block from the MoI apparatus. A hole was drilled in the front of the car for the string connected over the pulley to the hanging masses. Tape was used since the toy car multipacks contained many different types of cars, so taping the load was the only method that would work for all types of cars. Instead of the pulley-photogate system to measure the car's acceleration, the car was timed from start to 60 cm down the track (before washers hit the floor) multiple times via the acoustic stopwatch app to get an average time, and students used $d = \frac{1}{2}at^2$ to determine the car's acceleration for each hanging mass.

It was here that supply chain issues became an issue: the HotWheels track was nowhere to be found online or in stores at anything other than three times the normal price on eBay. But the combination of the blue tape, the meter stick, and some spare cardboard served as an adequate track for the experiment as shown in Figure 4.

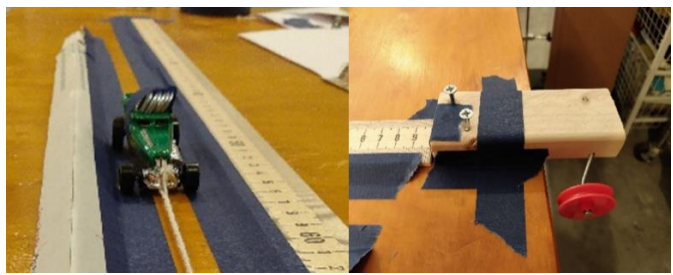

**Figure 4. Alternative for HotWheels track and pulley block from MoI apparatus for Newton's 2nd Law lab.**

Note that the track is now back in stock, and would only add $2 per student to the BOM. Since students subsequently reported that the blue tape setup required 15-30 minutes to setup alone, the track will be used when the kits are deployed in the Fall 2020 semester.

The collision in 2D experiment was then doable with the kit components already designed, all that was needed was a large piece of cardboard to put on the floor with carbon paper to mark ball hits from the steel ball alone, head-on collision of steel and glass balls, and glancing collision.

The centripetal force experiment was the only experiment that could not be easily replicated with low-cost home kit parts, so an alternative experiment that still met the activity learning goals was sought. The experiment chosen was one from Stonybrook University [3], and replicated all of the physics conceptual goals save the spring constant measurement aspect. A ¾" PVC pipe section cut from stock was used instead of the glass tube they used, and toy rubber ball with a hole drilled into it was used instead of rubber stopper so it could also be used as a plumb bob for the collisions in 2D experiment. After cutting the sections of PVC pipe, they had to be sanded so the sharp and rough edges would not fray the string nor hinder the smooth circular motion of the rubber ball. Note that in their first procedure, there is inherent bias in the slope from the small circle of rotation when slinging the ball that adds to the effective radius of the circular motion. Students discovered this for themselves from the larger percent difference with theory for this procedure, and the fact that the second procedure removes this bias from the slope and moves it to the y-intercept. Happily, the students learned how to modify experiments to remove bias through this exercise; perhaps this was an intension of the original experiment designers.

This still leaves a measurement of spring constant missing from the kit experiments. While a simple spring measurement experiment could be easily added to the kit via a single spring without raising the BOM significantly, due to the abbreviated summer semester it was not included. The spring experiment will be incorporated into the student guide and included in the kits when they are redeployed in the Fall 2020 semester.

There were a few other changes made for remote teaching of the lab due only to schedule limitations. First, an onsite equilibrium lab that used a force table with spring force gauges, a meter stick, hanging clips, and masses was replaced with online PHeT simulations [4] and a 2D force vector calculator from Desmos [5]. While low-cost force gauges are readily available, the addition of a large wooden disk to use as a force table would raise the shipping costs too much, and the online versions allowed all of the physics learning goals to be maintained. The physical meter stick equilibrium experiment could also easily be implemented with the existing kit parts using the included meter stick, washers, string instead of clips, and the wooden dowel from the MoI apparatus again as a clamp stand. Further, some instructors have occasionally replaced the meter stick equilibrium experiment with a commercial human arm model since many students are biology majors. It would be extremely simple to drill a hole in the meter stick, place the dowel through the hole, and use string and the pulley block, along with hanging masses, to implement a kit version of this experiment. Again, due only to time limitations, this experiment was not included in the short summer semester.

Figure 5 shows the kit as provided to students, minus the carbon paper and the 40" x 4" x 4" shipping box. Total BOM, including shipping boxes, worked out to just under $20 per student, well within the cost constraints decided upon in early discussions with university leadership. The components of the kits, along with mapping to the labs in the student guide and

modifications made to the standard equipment are listed in Table 1. Equipment for the two additional labs discussed above, which were not completed during the Summer 2020 session due to time constraints, are included on this list.

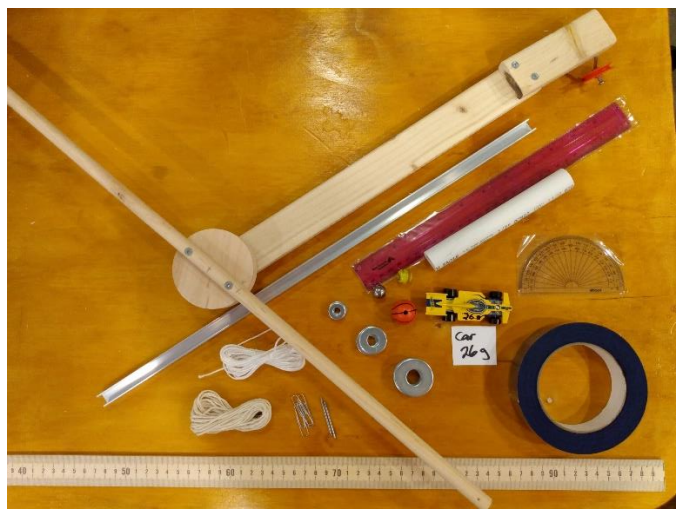

**Figure 5. Custom home experiment kits (minus carbon paper and shipping box).**

Prior to deployment of the kits, a sample kit was assembled and all experiments tested by the instructor so that a custom student guide for the kits could be made. The student guide is available from the authors' upon request. Next, the instructor's daughter, an undergraduate student herself, beta tested the kits and instructions, with the result that a few additional clarifications were added to the guides to ensure robust usage of the kits.

While a handful of students were more than 50 miles from the university, thus requiring shipping of the kits to them, most students lived near enough that they could pick the kits up. Following discussions with university leadership and facilities management, a protocol for pick-up that met their requirements and was satisfactory to the instructor was established. This greatly reduced shipping costs, which otherwise add 20% to the total cost. For the students that required shipping, only one box was damaged in shipment (breaking the MoI apparatus), and a replacement was quickly packed and sent in time for the start of the semester. The only remaining wrinkle, as also mentioned in Fox et al. was that an international student living in Egypt signed up for the course at the last minute. Given there was no way to get a kit to him, a plan was made to provide data from a lab partner. The student subsequently decided to wait and take the course in the fall so he could be onsite and do experiments hands-on.

**Table 1: Kit Components Mapped to Lab Activities**

| Kit contents | Lab Used | Notes |
|---|---|---|
| plastic protractor | measuring g (pendulum) | |
| 3 paper clips | all except collisions | |
| 2 small nails | moment of inertia | |
| 6 large washers (AGB) | all except collisions | find average mass and report to students |
| 6 med washers (ACB) | all except collisions | find average mass and report to students |
| 6 small washers | all except collisions | find average mass and report to students |
| nylon string (2m) | centripetal force, collisions | |
| cotton string (3m) | all except centripetal force and collisions | |
| HotWheels™ Car and mass spec | Newton's 2nd | drill hole in car front, provide mass of each to students |
| steel ball | measuring g, collisions | find average mass and diameter and report to students |
| glass ball (marble) | collisions | find average mass and diameter and report to students |
| rubber ball | centripetal force, collisions | drill hole through center, find average mass and report to students |
| aluminum u-channel rail (24" x 1/2") | measuring g, collisions | cut from stock |
| meter stick | all | drill 1/2" hole for equilbrium* |
| flexible Ruler with fuller | measuring g, collisions | drill 4 holes in corners |
| 3/4" PVC tube | centripetal force | sand ends |
| *moment of inertia (MoI)* apparatus (1/2" dowel, plywood drum, 2" lazy-susan base, 20" length of 1x2 wood with removable pulley block) | measuring g, Newton's 2nd, equilibrium* | build from pieces; bend nail into zigzag shape for holding pulley |
| blue painter's tape | all | |
| 3 sheets carbon paper | measuring g, collisions | |
| two light springs* | Hooke's law* | |
| **Student Supplied** | | |
| cell phone with PHYPHOX app | all | acoustic stopwatch app |
| Phillips screwdriver | Newton's 2nd | remove pulley block from MoI apparatus |
| cardboard | measuring g, collisions | |

*used for additional labs (Hooke's Law, full equilibrium/torque) for longer semester

## V. MATCHING LEARNING GOALS WITH THE KITS AND ONLINE IMPLEMENTATION

By developing at-home experiments that replicated the specific objectives of each of the traditional in-person labs we were able to address course learning goals LG-1 through LG-10, as outlined in Section II, at least as well as we had been doing in the onsite labs. In order to ensure that the remote experience closely reflected the onsite experience, it was decided that lab sessions would be synchronous, but also recorded for students with issues such as changes in work, family illness, and so on. As before, labs alternated between simulation ('discovery') and experiment labs. Students were required to set up the experiments prior to the start of the synchronous session, and each student had to do all standard experiments in the guide and submit their own spreadsheet (a template was provided in first few experiments only as a model) and students also had to do an online questionnaire, the latter included specifically as a way to check on critical learning goals. Since the semester was condensed and two labs per week were done, the normal writing of lab reports for each lab was suspended, and the scientific communication and documentation goals were met via the group project final report. To keep the collaboration aspect, students were allowed to do breakout rooms on Zoom during synchronous lab sessions.

In a traditional semester on campus, the group project assignment allows students to form groups and choose a specific measurement to make, which is often tied to their career goals. A small budget is available to students to order parts needed to perform the measurement. But students would not be able to collaborate in the same manner during the pandemic lockdown. Since all students possessed the experiment kits however, it was decided that group projects would involve each student choosing an extension to the standard experiments to design and perform. Students would then form groups who were extending the same standard experiment, and each had to do a different extension.

The student guides had suggested extensions, but many students came up with their own ideas. As done previously, the final report was completed in a provided template like a journal paper and titled as a letter to the editor. Tables and figures had to be professionally done, well-labeled, and with proper units. One week before it was due, students submitted drafts that were reviewed and edited by both the instructor and by multiple student peer reviewers from other groups; peer reviews were graded for helpfulness and professionalism.

Finally, and to continue to leverage the fact that all students had the kits available to them 24/7, students were offered extra credit points to do further extensions to standard experiments with the kits that were different from the extensions they did for the group project.

## VI. INITIAL OBSERVATIONS AND TECHNICAL PERFORMANCE

The first experiment lab (measuring g) is notoriously long, and it took 4+ hours for most students to complete. This was still shorter than the combined lab plus report preparation previously in the onsite version, however the student tolerance for this length of assignment was low. After the first experiment lab, students requested a video of the setup procedure in addition to the student guide; the instructor provided videos subsequently, and as noted in Fox et al., and known to all who have produced online videos, it often takes 7-8 times the running time of a video to produce it well. For this reason, setup videos were kept very short, but also to ensure students still had to figure things out.

Importantly, no "tuning out" was ever observed; students remained busy and industrious during the entire lab session until they completed their work. Many students also redid experiments and resubmitted their spreadsheets with improved data/analysis after discussing the lab with the instructor in online office hour sessions. And as expected, students used the extra credit opportunity to make up for poor grades on individual labs, and happily, often chose a different experiment to extend. Some even submitted multiple additional extensions to improve their overall score.

The technical performance of the measuring g lab met expectations, although students' values of g were further off from actual g than expected from Mentzer's paper. It is suspected that the shipping boxes as targets may have been not been level throughout the student's trials, thereby causing greater variation. But note that some students did indeed get within a few percent of the actual value, so perhaps with some additional guidance to keep the boxes level, the results will improve. The complexity of the instantaneous velocity (Merton method) calculation throws many students off, just as in the onsite picket fence free fall experiment, but is still considered an important learning goal for the lab. Some students complained that this lab took them over 5 hours to complete, and it was the least favorite of many according to the survey. After the students had completed the experiment, some modifications were made to the student guide in order to reduce experimental errors.

The technical performance of the labs as seen in students' submissions are summarized in Table 2. Results indicate the clear advantage of automatic timing in onsite labs (photogate used in both Measuring g and Newton's 2nd Law labs). While not necessary to learn physics, accuracy in results did appear to lead to greater student satisfaction from general observations over the years and also from the remote teaching experience reported here. However lower accuracy may have been an additional motivator for students to repeat experiments on their own, so perhaps this is also an added benefit.

If there is any global recommendation to improve technical performance, it would be to look for a low-cost timing/automation device to add to the kits, or to find a way for them to automatically time events, such as designing noisy barriers for the cars to hit at the finish line of the Newton's 2nd Law lab.

**Table 4. Comparison of technical performance of onsite vs. home kit experiments.**

| Experiment Lab | % of Students with >90% Accuracy | |
|---|---|---|
| | Onsite | Kit |
| Measuring g* | 65% | 33% |
| Newton's 2nd Law | 95% | 21% |
| Centripetal Force | N/A | 38% |
| Moment of Inertia | 40% | 40% |
| Collisions in 2D | 75% | 75% |

VII. STUDENT ATTITUDE SURVEY AND RESULTS

At the end of the Summer 2020 semester students completed a survey to explore their attitudes about the use of the kits and what they felt they had learned as a result of having them, as compared to pure simulations and/or instructor provided data. Survey questions included both Likert-scale and free response prompts, as shown in Table 3. Average scores and standard deviations, with all but one student in the course responding, are reported. We also include a mapping of the survey questions to the related learning goals, as numbered in Section II. Although the instrument has not been validated, it serves as a coarse measure of the efficacy of this instructional method.

We can see that on average, students agree-strongly agree to most of the questions. Students do seem to perceive a benefit from working with the kits that they do not feel they would have gotten from only working with simulations or instructor-provided data sets. Student responses to questions SU08 and SU09 suggest an increased confidence in performing laboratory work; this sentiment was echoed in several student comments in response to SU22. Students reported learned skills in line with the course learning goals, including creating plots (SU30) and technical writing (SU34).

More significantly, while Fox et al. reported that "a majority of the students felt that the remote classes were the same or worse than in-person labs," in the case reported here, the opposite was true: a majority of the students felt that the remote labs were equal to the in-person labs (SU23). And while many instructors reported in Fox et al. that student agency and/or engagement suffered after going remote, in the case reported here students were seen to be continuously engaged, both in the synchronous lab sessions and in the group projects. And students overwhelmingly agreed that 'freeloaders' (students who contributed little to group projects) were eliminated by the fact that all students had to perform all of the experiments individually and also extend those results individually in their group project assignment (SU35).

In addition to the Likert-scale responses, student responses to SU22: *What three positive things do you feel you specifically learned from the kit experiments that you would not have learned from simulations or instructor-provided data*, revealed much about the perceived benefits of working with the kit experiments. Student responses were binned independently by the authors along a number of different categories, which were then mapped to related course learning goals (Figure 6). Of the 23 students responding, 14 students reported learning an increased ability to troubleshoot experimental set-ups and check their results as a consequence of working with the lab kits. While not a specific learning goal, 9 students indicated that they enjoyed the hands-on experience, which they would not have gotten with other approaches to remote labs. Many students also reported an increased understanding of the role of uncertainty in experiment analysis, and an increased ability to identify and quantify sources of uncertainty.

**Table 3: End of Semester Survey Questions with Course Average (N = 23)**

| 1 - Strongly Disagree   2 - Disagree   3 - Neutral   4 - Agree   5 - Strongly Agree | Average | StDev | Mapped Goal |
|---|---|---|---|
| SU01-The kit experiments and extensions helped me learn experiment design in ways not possible with simulations. | 4.0 | 1.1 | LG-1 |
| SU02-The kit experiments and extensions helped me learn analysis of real-world data in ways not possible with simulations. | 4.0 | 0.9 | LG-1 |
| SU03-The kit experiments and extensions helped me develop intuition with physics equations in ways not possible with simulations or homework problems. | 4.0 | 1.0 | LG-8 |
| SU04-The kit experiments helped me understand the nature of scientific measurements (repeatability, uncertainty, bias, precision) in ways not possible with simulations. | 4.2 | 0.7 | LG-3 |
| SU05-The kit experiments helped me understand the nature of scientific measurements (repeatability, uncertainty, bias, precision) in ways not possible with instructor-provided data. | 3.8 | 1.0 | LG-3 |
| SU06-The kit experiments allowed me to learn how to determine sources of uncertainty in ways not possible with simulations or instructor-provided data. | 4.1 | 0.9 | LG-4 |
| SU07-Running the experiments myself taught me that it's OK to get inconsistent results if that is what the data analysis revealed. | 4.5 | 0.5 | LG-6 |
| SU08-As a result of running all experiments by myself, I feel that I am a better experimentalist now. | 4.2 | 0.6 | LG-5 |
| SU09-As a result of running all experiments by myself, I am more confident about analyzing other's experimental results to identify flaws in procedures and/or analysis. | 3.9 | 0.9 | LG-10 |
| SU10- As a result of running all experiments by myself, I feel I now have better troubleshooting skills for real-world experiment situations. | 4.2 | 0.8 | LG-5 |
| SU11- The kit experiments allowed me to develop critical thinking for laboratory experiments in ways that would not be possible with simulations or instructor-provided data. | 3.9 | 1.2 | LG-6 |
| SU12-Using the kits for the extensions from the standard experiments helped me understand how modeling is used to explore equations and physics concepts. | 4.0 | 0.9 | LG-7 |
| SU13-My experiences with the lab kits will help me apply physics concepts to novel situations. | 3.8 | 0.9 | LG-8 |
| SU14-The kit extensions as the group project or extra credit helped me to learn how to set scientific research goals. | 4.0 | 0.9 | LG-12 |
| SU15-The kit extensions as the group project or extra credit helped me learn how to design, evaluate feasibility, and improve custom scientific experiments. | 4.2 | 0.8 | LG-13 |
| SU16-The hands-on experimentation experience I got from PHY101L with the home kits will help me do better in other science laboratory courses. | 3.9 | 0.9 | |
| SU17-I took advantage of having the experimental equipment at home to redo my experiment on at least one occasion. | 3.9 | 1.1 | |
| SU18-Of all the kit experiment labs, my favorite was: | | | |
| SU19-Why was it your favorite? | | | |
| SU20-Of all the kit experiment labs, my least favorite was: | | | |
| SU21-Why was it your least favorite? | | | |
| SU22-What three positive things do you feel you specifically learned from the kit experiments that you would not have learned from simulations or instructor-provided data. | | | |
| SU23-I feel that I learned as much through this online experience as I would have in a face-to-face lab. | 3.1 | 1.3 | |
| SU24-Having to do the experiments by myself at home was harder than in a group onsite in the lab. | 4.5 | 0.9 | |
| SU25-We use this statement to discard the survey of people who are not reading the questions. Please select agree (not strongly agree) for this question to preserve your answers. | 4.0 | 0.0 | |
| SU26-Having templated worksheets took all the creativity out of worksheet design; I would rather design my own worksheets. | 1.7 | 1.1 | |
| SU27-The extension options for the experiments allowed me to learn how to create theoretical models associated with, and at least partially agree with, the experimental data via the worksheets. | 4.0 | 1.0 | LG-1 |
| SU28-The practice with uncertainty in each experiment lab via the worksheets allowed me to hone my skills and more quickly and accurately identify and calculate uncertainty in future experiments. | 3.8 | 1.2 | LG-4 |
| SU29-The worksheets and associated analysis helped me learn how to analyze and visualize data in a variety of ways not possible with short answer-type lab reports. | 4.3 | 1.1 | LG-9 |
| SU30-The plots required in the worksheets helped me learn how to create publishable charts for research papers. | 4.3 | 0.6 | LG-11 |
| SU31-My spreadsheet skills are significantly improved as a result of this course. | 4.5 | 1.0 | LG-2 |
| SU32-The group project report helped me in a significant way to learn how to communicate scientific results. | 3.5 | 1.3 | LG-11 |
| SU33-The instructor and peer-review of our group project report were instrumental to having a publishable-quality report. | 4.0 | 0.9 | LG-11 |
| SU34-As a result of feedback I got from the instructor and peer-reviewers, I will be a better technical author in the future. | 4.1 | 1.0 | LG-11 |
| SU35-The fact that all group project team members had to do the kit-based experiments by themselves, as well as their own extension, prevented the situation where some team members do little to no work on the group project report. | 4.1 | 1.1 | |
| SU36-By comparing our individual results from the standard kit experiment in our group project, we learned even more about the nature of scientific measurements (repeatability, uncertainty, bias, precision). | 4.0 | 1.0 | LG-3 |

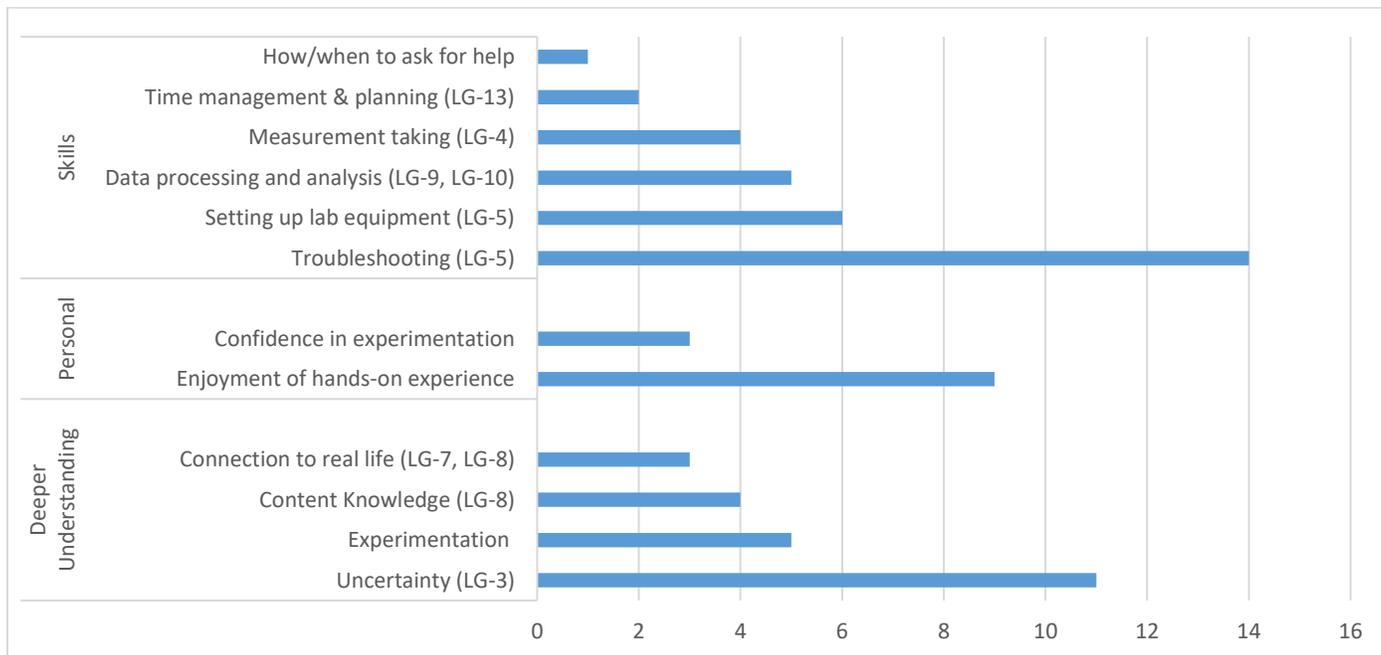

**Figure 6: Binning of student responses to end-of-semester survey question SU22. Categories were then mapped to related course learning goals, as noted in parentheses.**

## VIII. SUMMARY

We present in this paper an inexpensive custom home experiment kit for first semester physics labs. The construction and deployment of the kits was completed during the abbreviated Summer 2020 semester at Oglethorpe University, which was conducted entirely remotely due to the COVID-19 pandemic. The kit replicates five common lab experiments, with the possibility of an additional sixth experiment. Instructions for building the kits are described; an associated student guide, referenced in the text, is available upon request from the authors.

The kits were designed to address the original onsite activity learning goals, and the remote course was administered so as to meet the overall course learning goals. A study of remote physics lab implementation in response to the COVID-19 pandemic by Fox et al. concluded that "many critical learning goals are hard, if not impossible, to meet in a fully remote class." This was largely disproven in the present case where a combination of home experiment kits customized to match the onsite labs, along with a group project designed to leverage the home kits and specifically engineered for the remote scenario worked extremely well at matching the critical learning goals of the onsite labs.